\documentclass[aps,prd,groupedaddress,nofootinbib]{revtex4-2}
\usepackage{graphicx}
\usepackage{slashed}
\usepackage{xcolor}

\newcommand{\be}{\begin{equation}}
\newcommand{\ee}{\end{equation}}
\newcommand{\ba}{\begin{eqnarray}}
\newcommand{\ea}{\end{eqnarray}}
\newcommand{\ban}{\begin{eqnarray*}} 
\newcommand{\ean}{\end{eqnarray*}}

\def\d{\partial} 
\def\b{{\mathbf b}}
\def\p{{\mathbf p}}

\def\k{{\mathbf k}}
\def\x{{\mathbf x}}  
\def\y{{\mathbf y}}
\def\r{{\mathbf r}}
\def\z{{\mathbf z}}

\def\v{{\mathbf v}}

\def\K{{\mathbf K}}

\def\P{{\mathbf P}}

\begin{document}


\title{Probing quark transverse momentum distributions in the Color Glass Condensate: quark-gluon dijets in Deep Inelastic Scattering at next-to-eikonal accuracy}


\author{Tolga Altinoluk}
\affiliation{ Theoretical Physics Division, National Centre for Nuclear Research,
Pasteura 7, Warsaw 02-093, Poland}
\author{N\'estor Armesto}
\affiliation{Departamento de F\'{\i}sica de Part\'{\i}culas and IGFAE, Universidade de Santiago de Compostela, 15782 Santiago de Compostela, Galicia, Spain}
\author{Guillaume Beuf}
\affiliation{Theoretical Physics Division, National Centre for Nuclear Research,
Pasteura 7, Warsaw 02-093, Poland }


\date{\today}

\begin{abstract}
We study the production,  in Deep Inelastic Scattering at high energy, of a quark-gluon dijet  induced by $t$-channel quark exchange with the target, which goes beyond the eikonal approximation. Throughout this study we follow the Color Glass Condensate approach, keep full dependence on the quark mass and consider the target to be unpolarized. We focus on the correlation limit in which the produced jets fly almost back-to-back and find a factorized expression for the cross section, for both longitudinal and transverse photons, involving the unpolarized quark transverse momentum dependent distribution.
Quark-gluon dijets can be distinguished from other types of dijets at least in the heavy quark case, thanks to heavy flavor tagging. In that case, the dominant background is expected to come from the eikonal production of a quark-antiquark-gluon system. We propose experimental cuts to suppress that background contribution, making the process of quark-gluon dijet production in Deep Inelastic Scattering a new method to probe the quark transverse momentum dependent distribution at the Electron Ion Collider.

\end{abstract}


\maketitle

\section{Introduction}

Understanding the high-energy behavior of Quantum Chromodynamics (QCD) is one of the key missing aspects in our knowledge of the strong interaction. Such understanding, intrinsically linked with that of the structure of hadrons in the region where their partons carry a small fraction $x$ of the hadron momentum, is essential for precise predictions of  the production of Standard Model (SM) particles at hadronic colliders. Besides, QCD processes are the largest background in the determination of SM parameters and searches for physics beyond the SM.

On general grounds, it is expected that at high energies QCD enters a regime where the linear approximation to QCD radiation breaks and non-linear effects dominate~\cite{Kovchegov:2012mbw}. Such non-linear effects, leading to the saturation of parton densities, are enhanced at small values of $x$ and for nuclei. The Color Glass Condensate (CGC) effective field theory~\cite{Gelis:2010nm} offers a weak coupling but non perturbative realization of parton saturation.

ln recent years it has been realised that the objects describing hadrons in CGC calculations (namely, ensembles of Wilson lines averaged over the color configurations of the hadron) can be related, considering a certain kinematic situation called correlation limit, to the parton distributions discussed in the frame of transverse momentum dependent (TMD) factorization~\cite{Collins:2011zzd,Boussarie:2023izj}. Such relation for dijet ~\cite{Dominguez:2011wm,Dominguez:2012ad,Marquet:2016cgx,Petreska:2018cbf,Marquet:2017xwy} and trijet \cite{Altinoluk:2018uax,Altinoluk:2018byz,Altinoluk:2020qet} production  have been established for the gluon TMD parton distribution functions at leading order (LO) in the coupling constant at leading twist.\footnote{The sea quark contribution to the quark TMD
stemming from gluon splitting
was also studied within the CGC in the eikonal limit~\cite{Marquet:2009ca,Hatta:2022lzj}.} Recently, this relation is studied at next-to-leading order (NLO) \cite{Taels:2022tza,Caucal:2022ulg} and also at LO but beyond the kinematic leading twist approximation \cite{Kotko:2015ura,vanHameren:2016ftb,Altinoluk:2019fui,Altinoluk:2019wyu,Boussarie:2020vzf,Fujii:2020bkl,Altinoluk:2021ygv,Boussarie:2021ybe} for dijet production. 

On the other hand, in CGC calculations and, in general, in high-energy QCD the eikonal approximation is used, which results in discarding terms sub-leading in energy.
 In recent years, there have been a lot of efforts to go beyond the eikonal approximation and account for the sub-eikonal contributions \cite{Altinoluk:2014oxa,Altinoluk:2015gia,Altinoluk:2015xuy,Agostini:2019avp,Agostini:2019hkj,Agostini:2022ctk,Agostini:2022oge,Altinoluk:2020oyd,Altinoluk:2021lvu,Altinoluk:2022jkk,Kovchegov:2015pbl,Kovchegov:2016zex,Kovchegov:2016weo,Kovchegov:2017jxc,Kovchegov:2017lsr,Kovchegov:2018znm,Cougoulic:2019aja,Kovchegov:2020hgb,Cougoulic:2020tbc,Adamiak:2021ppq,Kovchegov:2021lvz,Kovchegov:2021iyc,Cougoulic:2022gbk,Kovchegov:2022kyy,Chirilli:2018kkw,Chirilli:2021lif,Balitsky:2015qba,Balitsky:2016dgz,Balitsky:2017flc,Boussarie:2020fpb,Boussarie:2021wkn,Jalilian-Marian:2017ttv,Jalilian-Marian:2018iui,Jalilian-Marian:2019kaf} in various processes. It has been shown that in some observables, such as the ones sensitive to spin, see e.g.~\cite{Cougoulic:2022gbk}, the terms sub-leading in energy dominate the cross section. 
 Computations at sub-eikonal accuracy require the consideration of quarks, and not only gluons, exchanged in the $t$-channel (see e.g.  \cite{Kovchegov:2015pbl,Chirilli:2018kkw}).

The study of hadron and nuclear three-dimensional structure and the search for the non-linear regime of QCD are, together with the understanding of nucleon spin, the physics pillars of the Electron Ion Collider (EIC)~\cite{Accardi:2012qut,AbdulKhalek:2021gbh}. The very large luminosity, together with modern large acceptance detectors, offer unprecedented opportunities for QCD studies. The center-of-mass energy $\sqrt{s_{NN}}\simeq 40-140$ GeV/nucleon, similar to those at the Relativistic Heavy Ion Collider at BNL and lower than at the Large Hadron Collider at CERN or the Hadron-Electron Ring Accelerator at DESY, together with the spin program, require a careful consideration of sub-eikonal effects.

\begin{figure}[htb]
 \includegraphics[width=8.6cm]{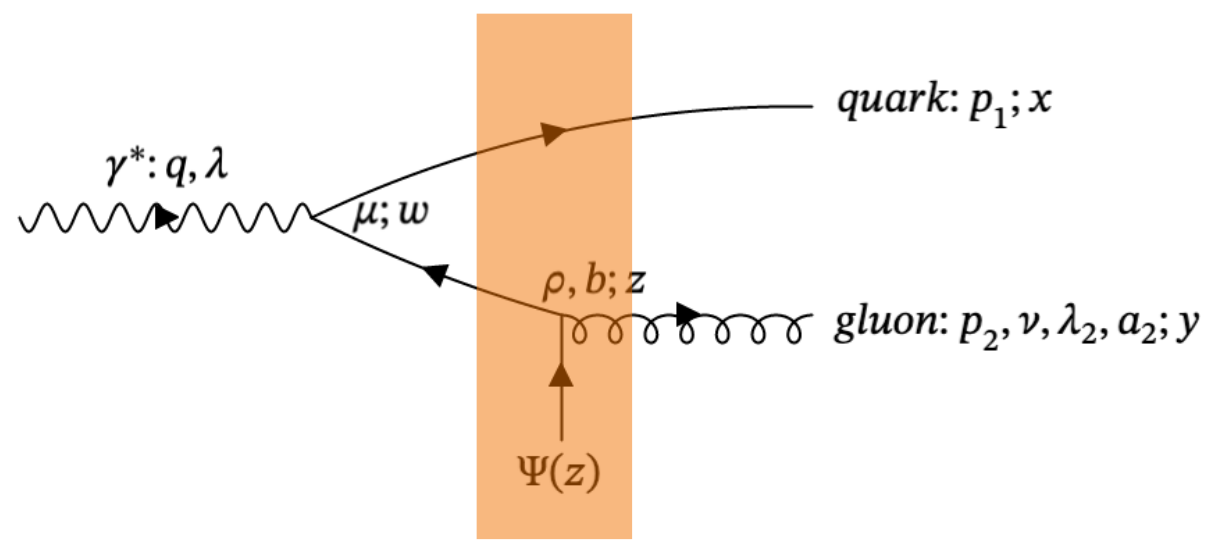}
 \includegraphics[width=8.6cm]{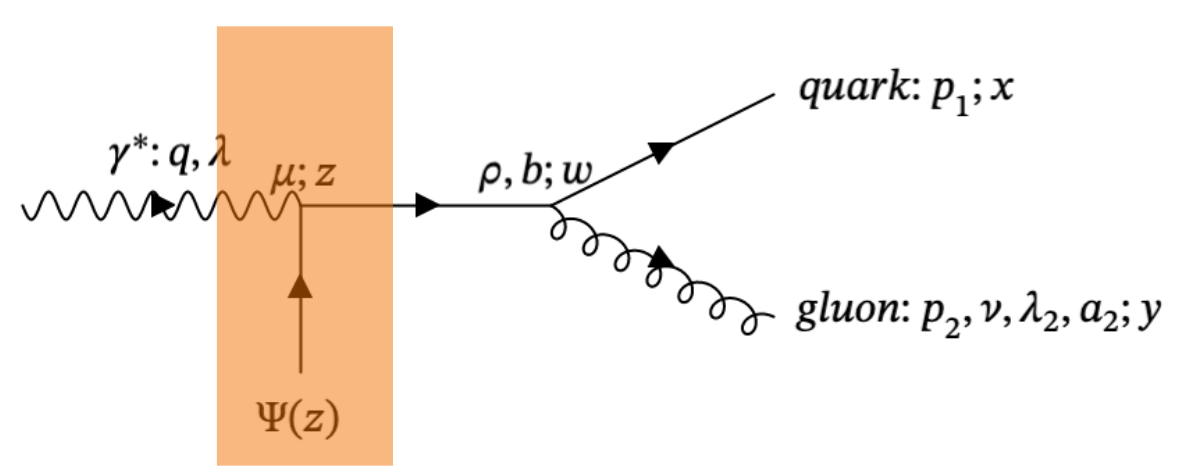}
\caption{Diagrams with photon splitting before (left) or interacting inside (right) the target. Momenta, color and polarization indices, and coordinates separated by a semicolon, are indicated. The target hadron is shown as an orange band.}
\label{fig:diag}
\end{figure}

In this context, here we establish for the first time the relation at small $x$ between CGC and TMD calculations including quark $t$-channel exchanges that are subdominant in energy and thus can be considered sub-eikonal. We examine a process determined by the $t$-channel exchange of a quark: the LO production of a quark-gluon dijet in Deep Inelastic Scattering (DIS) on an unpolarized target. We first compute the scattering amplitudes for transverse and longitudinal virtual photons within the CGC framework. Calculations are done keeping the quark mass. We then go to the correlation limit and compute the cross section. In this limit, the CGC average of the Wilson lines obtained for this process, characterizing the target, can be related to the quark TMD distribution in the hadron target at small $x$. The cross section shows explicit suppression with the energy compared to gluon exchange.  Finally, we discuss how to disentangle this mechanism for $qg$ dijet production from the contributions that do not involve quark TMD distributions.


\section{Scattering amplitudes for quark-gluon dijet production at next-to-eikonal accuracy in DIS}

With the previously mentioned motivation, we are interested in studying the production of a quark and a gluon (to be interpreted as separate jets) from the interaction of a virtual photon with a nucleus target. More precisely, we focus on the contribution from a single $t$-channel quark exchange at the amplitude level which is sub-eikonal. But we resum multiple interactions with the gluon background field. The two relevant diagrams in this case correspond to two different orderings between the photon and radiated gluon vertices along the quark line as shown on Fig.~\ref{fig:diag}. At high scattering energies, due to the large boost of the target, the quark background field is effectively localized in a parametrically small support along the $x^+$ direction. This forces one of the two vertices to be inside the small support of the target. By contrast, the other vertex has to be outside of the target in order to avoid further power suppression at high energy. The leading power contributions are shown in Fig.~\ref{fig:diag}. In the first one (called {\it before} contribution), the photon splits into quark-antiquark before the target like in the standard dipole picture of DIS \cite{Nikolaev:1990ja} and then the antiquark is converted to a gluon via interaction with the $t$-channel quark.  In the second one (called {\it in} contribution), the photon is converted to a quark which then radiates a gluon after exiting the target. At $S$-matrix level, these two contributions can be written as\footnote{Throughout this paper, we use the light cone basis for momenta $p=(p^+,p^-,{\bf p})$ with $p^\pm=(p^0\pm p^3)/\sqrt{2}$. Moreover, in what follows $m$ is the quark mass and $Q^2=-q^2$ is the photon virtuality, and subindices in integrals denote integration variables, i.e., $\int_x=\int dx$, $\int_{\bf x}=\int d^2{\bf x}$.}   
%

\ba
S^{\rm bef}_{\gamma\to q_1g_2}&=&\lim_{x^+, y^+\to+\infty}\int_{\x,\y}\int_{x^-,y^-} \!\!e^{ip_1\cdot x}\, {\bar u}(1)\gamma^+ \, e^{ip_2\cdot y}\epsilon^{\lambda_2}_{\nu}(p_2)^*(-2p_2^+) 
\nonumber\\
&&\times 
\int _{w,z} e^{-iq\cdot w} \epsilon^{\lambda}_{\mu}(q)\, G_F^{\nu\rho}(y,z)_{a_2 b}
S_F(x,w)(- ie e_f)\gamma^{\mu} S_F(w,z)
(-ig)\gamma_{\rho}t^b\Psi(z),
\label{bef-def}\\
S^{\rm in}_{\gamma\to q_1g_2}&=&\lim_{x^+, y^+\to+\infty}\int_{\x,\y}\int_{x^-,y^-} \!\!e^{ip_1\cdot x}\, {\bar u}(1)\gamma^+ \, e^{ip_2\cdot y}\epsilon^{\lambda_2}_{\nu}(p_2)^*(-2p_2^+)
\nonumber\\
&&\times 
\int _{w,z} e^{-iq\cdot z} \epsilon^{\lambda}_{\mu}(q)\, G_{F,0}^{\nu\rho}(y,w)_{a_2 b}
S_{F,0}(x,w) (-ig)\gamma_{\rho}t^b S_F(w,z)
(- ie e_f)\gamma^{\mu}\Psi(z),
\label{in-def}
\ea
for arbitrary polarization $\lambda$ of the photon. Here, $\Psi(z)$ is the quark background field,  $S_F$ and $G_F$ are the Feynman quark and gluon propagators in the gluon background field of the target at eikonal accuracy, whereas $S_{F,0}$ and $G_{F,0}$ are the corresponding vacuum propagators. The scattering amplitude is defined from the $S$-matrix element as follows:
\ba
S_{\gamma_{T,L}\to q_1g_2}&=& (2q^+)\, 2\pi \delta(p_1^+\!+\!p_2^+\!-\!q^+)\, i {\cal M}_{\gamma_{T,L}\to q_1g_2}\; .
\label{def:amp}
\ea
%


%
In order to define the power counting associated with the eikonal approximation and to classify sub-eikonal corrections, we follow the approach from Refs.~\cite{Altinoluk:2020oyd, Altinoluk:2021lvu, Altinoluk:2022jkk}, based on scaling properties under a large boost of the target along $x^-$. In particular, components of vector and tensor quantities associated only with the target have simple scaling properties in light-cone coordinates, determined by their number of upper and lower $+$ and $-$ indices. In the case of a current associated with the target (for example a color, or flavor or baryon number current), the components should scale as
\be
J^{-}(z) \propto \gamma_t\, , \;\;\; 
J^{j}(z) \propto (\gamma_t)^0\, , \;\;\; 
J^{+}(z) \propto (\gamma_t)^{-1}
\label{target_current_scaling}
\, ,
\ee
under a large boost of the target along the $x^-$ direction, with boost factor $\gamma_t$.

In order to setup our calculation, we also need to specify the scaling behavior of the quark background field $\Psi(z)$ of the target. It is convenient to introduce the projections
\ba
\Psi^{(-)}(z)
&\equiv &
\frac{\gamma^+\gamma^-}{2}\, \Psi(z),\nonumber\\
\Psi^{(+)}(z)
&\equiv &
\frac{\gamma^-\gamma^+}{2}\, \Psi(z)
\label{psi_proj}
\, ,
\ea
respectively, on the so-called good (or partonic) components of $\Psi(z)$ of the target and on the bad (or non-partonic) components of $\Psi(z)$ of the target. 
 Then, currents constructed as bilinears of  $\Psi(z)$ have components depending on these two types of components as
\ba
\overline{\Psi}(z)\,  \gamma^-\, \Psi(z) &=&  \overline{\Psi^{(-)}}(z)\,  \gamma^-\, \Psi^{(-)}(z) ,\nonumber\\
\overline{\Psi}(z)\,  \gamma^j\, \Psi(z) &=&  \overline{\Psi^{(-)}}(z)\,  \gamma^j\, \Psi^{(+)}(z) 
+\overline{\Psi^{(+)}}(z)\,  \gamma^j\, \Psi^{(-)}(z) , \nonumber\\
\overline{\Psi}(z)\,  \gamma^+\, \Psi(z) &=&  \overline{\Psi^{(+)}}(z)\,  \gamma^-\, \Psi^{(+)}(z) 
\, .
\ea
 Such currents associated with the target have to follow the scaling behavior (\ref{target_current_scaling}), so that the components of the quark background field scale as
\ba
\Psi^{(-)}(z)
&\propto & (\gamma_t)^{\frac{1}{2}},
\nonumber\\
\Psi^{(+)}(z)
&\propto & (\gamma_t)^{-\frac{1}{2}},
\ea
under a large boost of the target. 
In our calculation at next-to-eikonal (NEik) accuracy, only the components $\Psi^{(-)}(z)$ matter in  Eqs.~(\ref{bef-def}) and (\ref{in-def}), whereas the components $\Psi^{(+)}(z)$ would contribute only to corrections which are further suppressed at high energy (next-to-next-to-eikonal (NNEik) and beyond).

Therefore, at $S$-matrix level, both contributions (\ref{bef-def}) and (\ref{in-def}) have an overall scaling as $(\gamma_t)^{-\frac{1}{2}}$ under a large boost of the target along the $x^-$ direction, resulting from the enhancement as  $(\gamma_t)^{\frac{1}{2}}$ of $\Psi^{(-)}(z)$  and from the suppression as $(\gamma_t)^{-1}$ due to the integration in $z^+$ over the Lorentz contracted support of $\Psi(z)$, representing the width of the target (see Refs.~\cite{Altinoluk:2020oyd, Altinoluk:2021lvu, Altinoluk:2022jkk}).
Hence, at cross section level, a suppression as  $(\gamma_t)^{-1}$ is obtained, corresponding indeed to NEik order.
 

 
As a remark, from Eqs.~(\ref{bef-def}) and (\ref{in-def}), one obtains as well contributions induced by the instantaneous part of the intermediate quark propagator $S_F(w,z)$ (see Eq. (17) in Ref. 
\cite{Altinoluk:2020oyd}) . In that case, the photon annihilation and the gluon emission happens simultaneously (meaning $w^+=z^+$). However, due to the Dirac structure of the instantaneous part of the quark propagator and to the light-cone gauge polarization vectors for QCD and QED, the enhanced components $\Psi^{(-)}(z)$ of the quark background field are projected out, and only the suppressed components $\Psi^{(+)}(z)$ survive. For that reason, the diagrams with simultaneous photon annihilation and gluon emission contribute only at NNEik accuracy at cross section level instead of NEik for the terms calculated in the present paper.

In order to evaluate the "in" contribution, corresponding to Eq. (\ref{in-def}) and represented on the right panel of Fig. \ref{fig:diag}, one needs the quark propagator from the point $y$ inside the target to the point $x$ after the target at eikonal accuracy. It is derived in Ref.~\cite{Altinoluk:2022jkk} and the final result reads
\ba
&&
{S}_F(x,y) \bigg|^{\rm{IA, }q}_{\rm{Eik.}}
=
\int \frac{d^2\p}{(2\pi)^2}\, \frac{dp^+}{(2\pi)}\frac{\theta(p^+)}{2p^+}\, 
e^{-i x\cdot {p}}\, 
(\slashed{{p}}+m)\;
 \mathcal{U}_F(x^+,y^+;\y)\;
\left[ 1 \!- \!   \frac{ \gamma^+ \gamma^i}{2p^+}\, i\, \overleftarrow{{\cal D}^F_{\y^i}}\right]\,
 \, e^{i y^- p^+}\,  e^{-i \y\cdot \p}
\label{IA_q_propag_Eik}
\, ,
\ea
where $p^-\equiv (\p^2+m^2)/(2p^+)$.  

The Wilson lines that account for the multiple scattering of the projectile partons off the target gluon field are defined in the standard way as 
\ba
\hspace{-0.5cm}
U_R(x^+,y^+;\z)\equiv {\cal P}_+\exp \left\{-ig\!\!\int_{y^+}^{x^+}\!\!\!\!\!\!dz^+ T_R\!\cdot\! {\cal A}^-(z^+,\z)\right\},
\ea
where ${\cal P}_+$ denotes ordering of color matrices along $z^+$ direction. The subscript $R$ denotes the representation of the $SU(N)$ generators, either fundamental ($R=F$) or adjoint ($R=A$), and $t^a=T_F^a$ and $T^a=T_A^a$. Moreover, we also introduce the following shorthand notation:
\ba
\hspace{-0.5cm}
U_R(\z)\equiv U_R(+\infty,-\infty;\z)\, . 
\ea
The covariant derivatives in representation $R$ are defined as 
\ba
\overrightarrow{\mathcal{D}^R_{z^{\mu}}} \equiv &\, \overrightarrow{\d_{z^{\mu}}} +ig\, T_R\!\cdot\!\mathcal{A}_{\mu}(z), \\
\overleftarrow{\mathcal{D}^R_{z^{\mu}}} \equiv &\, \overleftarrow{\d_{z^{\mu}}} -ig\, T_R\!\cdot\!\mathcal{A}_{\mu}(z) 
\, .
\ea

In a similar way, in order to evaluate the "before" contribution,  corresponding to Eq. (\ref{bef-def}) and represented on the left panel of Fig. \ref{fig:diag},   one needs the quark propagator from the point $y$ inside the target to the point $x$ before the target and the gluon propagator from the point $y$ inside the target to the point $x$ after the target. These can be calculated at eikonal accuracy \cite{Altinolukxyz} and  the final results read
\ba
&&
{S}_F(x,y) \bigg|^{\rm{IB, }q}_{\rm{Eik.}}
=
\int \frac{d^2\p}{(2\pi)^2}\, \frac{dp^+}{(2\pi)}\frac{\theta(p^+)}{2p^+}\, 
e^{i x\cdot {p}}\, 
(\slashed{{p}}-m)\; (-1)
 \mathcal{U}^\dagger_F(x^+,y^+;\y)\;
\left[ 1 \!+ \!   \frac{ \gamma^+ \gamma^i}{2p^+}\, i\, \overleftarrow{{\cal D}^F_{\y^i}}\right]\,
 \, e^{-i y^- p^+}\,  e^{i \y\cdot \p}
\label{BI_q_propag_Eik}
\, 
\ea
and 
\ba
G^{\mu\nu}_F(x,y)\bigg|^{{\rm IA,} g}_{\rm Eik.}
&=&
\int \frac{d^2\p}{(2\pi)^2}\, \frac{dp^+}{(2\pi)}\frac{\theta(p^+)}{2p^+}\, 
e^{-i x\cdot {p}}
\left[ -g^{\mu j}+\frac{\p^j}{p^+}g^{\mu +}\right]\, {\cal U}_A(+\infty,y^+;\y)
\nonumber\\
&&\times\, 
\left[ g^\nu_{\;\;  j}+\frac{g^{\nu +}}{p^+}\left(\p^j+i\, \overleftarrow{{\cal D}^A_{\y^j}}\right)
 \right] e^{iy^-p^+}e^{-i\y\cdot \p}\, . 
\label{gluon_IA}
\ea

Moreover, we also need the well known expression for the quark eikonal propagator from the point $y$ before the target to the point $x$ after the target which reads 
\ba
{S}_F(x,y) \bigg|^{\rm{BA, }q}_{\rm{Eik.}}&=&\int \frac{d^2\p}{(2\pi)^2}\, \frac{dp^+}{(2\pi)}\frac{\theta(p^+)}{2p^+}\, 
\int 
 \frac{d^2\k}{(2\pi)^2}\, \frac{dk^+}{(2\pi)}\frac{\theta(k^+)}{2k^+}\, 
\, e^{-i x\cdot {p}}\, e^{i y\cdot {k}}\, 2\pi \delta(p^+-k^+)\, (\slashed{{p}}+m)\gamma^+(\slashed{{k}}+m)\nonumber\\
&&\times\, \int d^2\z \; e^{-i\z\cdot(\p-\k)}\, {\cal U}_F(+\infty,-\infty;\z).
\ea 
%

%
%

Following the formalism developed in \cite{Altinoluk:2020oyd, Altinoluk:2021lvu, Altinoluk:2022jkk,Altinolukxyz}, and using the propagators given in Eqs. (\ref{IA_q_propag_Eik}), (\ref{BI_q_propag_Eik}) and (\ref{gluon_IA}),  Eqs. (\ref{bef-def}) and (\ref{in-def}) can be evaluated and lead to the following contributions to the scattering amplitudes\footnote{In the case of longitudinal polarization of the photon, the contribution with photon splitting inside the target (\ref{in-def}) is found to vanish at NEik accuracy. Indeed, in light-cone gauge, $\slashed{\epsilon}_L(q)\, \Psi(z)= ({Q}/{q^+})\,\gamma^+ \, \Psi(z)$, and the $\gamma^+$ projects the quark field onto its subleading components, see Eq. (\ref{psi_proj}).
} 
\ba
i{\cal M}^{\rm bef}_{\gamma_{T,L}\to q_1g_2}&=&\frac{i e e_fg}{2q^+}  \int_{\v,\z}e^{-i\v\cdot\p_1-i\z\cdot\p_2}\int\frac{d^2\K}{(2\pi)^2}\frac{e^{i(\v-\z)\cdot\K}}{\left[ \K^2+m^2+\frac{p_1^+p^+_2}{(q^+)^2}Q^2\right]}\nonumber\\
&&\times 
{\bar u}(1)\frac{\gamma^+\gamma^-}{2} \Gamma^{\rm bef}_{T,L}
\int_{z^+} U_A\left(+\infty,z^+;\z\right)_{a_2b}
U_F(\v)U_F^\dagger\left(z^+,-\infty;\z\right)t^b
\Psi
(z^+,\z)\, ,\label{bef_S_gen}\\
i{\cal M}^{\rm in}_{\gamma_T\to q_1g_2}&=& \frac{i e e_fg}{2q^+}\, 
\frac{1}{ \left[ \left( \p_1-\frac{p_1^+}{p_2^+}\p_2\right)^2+m^2\right]}
\int_{\z}e^{-i\z\cdot(\p_1+\p_2)}
{\bar u}(1)\frac{\gamma^+\gamma^-}{2} \Gamma^{\rm in}_T
\int_{z^+} 
t^{a_2}U_F\left(+\infty,z^+;\z\right)
  \Psi
  (z^+,\z)
\label{in_S_gen},
\ea
with 
\be
\Gamma^{\rm bef}_L=2\frac{p_1^+p_2^+}{(q^+)^2}Q{\varepsilon^{j*}_{\lambda_2}}\gamma^j
\, , \hspace{0.5cm}
\Gamma^{\rm bef}_T=\varepsilon^i_\lambda \, {\varepsilon^{j*}_{\lambda_2}}
\left\{\K^l\left[\left(\frac{p_1^+-p_2^+}{q^+}\right)\delta^{il}-\frac{[\gamma^i,\gamma^l]}{2}\right]+m\gamma^i\right\} \gamma^j
\, ,
\ee
\be
\Gamma^{\rm in}_T=
{\varepsilon^{l*}_{\lambda_2}} \varepsilon^j_\lambda
\left\{ \left[\p_1^i-\frac{p_1^+}{p_2^+}\p_2^i\right]\left[-\left(\frac{2p_1^++p_2^+}{p_2^+}\right)\delta^{il}+\frac{[\gamma^i,\gamma^l]}{2}\right] +m\gamma^l\right\}
\gamma^j
\, .
\ee
%


 \section{Production cross section in general dijet kinematics}
 The quark-gluon dijet production cross section in DIS for the longitudinal photon case at next-to-eikonal accuracy can be written in general dijet kinematics (without assuming the correlation limit) as 
\ba
&&
(2\pi)^6\, (2p_1^+)\, (2p_2^+)\, 
\frac{d\sigma^{\gamma_L\to q_1g_2}}{dp_1^+d^2\p_1dp_2^+d^2\p_2}
=
(2q^+) \, (2\pi) \delta(p_1^++p_2^+-q^+)\, \sum_{h_1,\lambda_2}\sum_{\rm col.} \big| {\cal M}^{\rm bef}_{\gamma_L\to q_1g_2}\big|^2\, .
\label{def:xsec_L}
\ea

%
Using Eq. (\ref{bef_S_gen}), one gets 
\ba
 \sum_{h_1,\lambda_2}\sum_{\rm col.} \big| {\cal M}^{\rm bef}_{\gamma_L\to q_1g_2} \big|^2&=& \frac{e_f^2\, \alpha_{\rm em}\, \alpha_s}{(2q^+)}
 16 Q^2\, z^3(1-z)^2
 \int_{\b,\b',\r, \r'}e^{-i\k\cdot(\b-\b')}\; e^{i\P\cdot(\r-\r')} \, 
 K_0(|\r|\bar Q)\, K_0(|\r'|\bar Q)\int_{z^+,z^{\prime +}}\
 \nonumber\\
 &&
 \times\, 
 \bigg\langle\Big[ t^{b'}{\cal U}_F(z^{\prime +}, -\infty; \b'+z\r')\, {\cal U}_F^\dagger(\b'-(1-z)\r')\, {\cal U}_F(\b-(1-z)\r)\, {\cal U}_F^{\dagger}(z^+,-\infty; b+z\r)t^b\Big]_{\beta'\beta}\nonumber\\
 &&
 \times\, 
 \Big[ {\cal U}^\dagger_A(+\infty, z^{\prime +}; \b'+z\r')\, {\cal U}_A(+\infty, z^+;\b+z\r)\Big]_{b'b}\; \bar\Psi_{\beta'}(z^{\prime +}, \b'+z\r')\gamma^-\Psi_\beta(z^+, \b+z\r)\bigg\rangle.
\ea
The quark-gluon dijet production cross section in DIS for the transverse photon case at next-to-eikonal accuracy can be written as 
\ba
(2\pi)^6\, (2p_1^+)\, (2p_2^+)\, 
\frac{d\sigma^{\gamma_T\to q_1g_2}}{dp_1^+d^2\p_1dp_2^+d^2\p_2}
= 2q^+ \, (2\pi)\, \delta(p_1^++p_2^+-q^+)\, \frac{1}{2}\sum_{\lambda, \lambda_2h_1}\sum_{ {\rm col.}}\big| {\cal M}^{\rm bef}_{\gamma_T\to q_1g_2} + {\cal M}^{\rm in}_{\gamma_T\to q_1g_2} \big|^2\, . 
\label{def:xsec_T}
\ea
Using Eqs. (\ref{bef_S_gen}) and (\ref{in_S_gen}), one obtains 
\ba
\frac{1}{2}\sum_{\lambda, \lambda_2h_1}\sum_{ {\rm col.}}\big|  {\cal M}^{\rm in}_{\gamma_T\to q_1g_2}\big|^2&=&
\frac{e_f^2\, \alpha_{\rm em}\, \alpha_s\, C_F}{\big[ \P^2+(1-z)^2m^2\big]^2}\, \frac{(4\pi)^2}{2q^+}
\, z\, \Big[ (1+z^2)\P^2+(1-z)^4m^2\Big]
\int d^2\z\int d^2\z' e^{-i\k\cdot(\z-\z')} \nonumber\\
&&\times 
\int_{z^+,z^{\prime +}}
\left\langle {\bar \Psi}(z^{\prime +},\z)\, \gamma^-\, {\cal U}_F^\dagger(+\infty, z^{\prime+}; \z')\, {\cal U}_F(+\infty,z^+; \z)\, \Psi(z^+, \z)\right\rangle,
\ea
\ba
&&
\frac{1}{2}\sum_{\lambda, \lambda_2h_1}\sum_{ {\rm col.}}\big| {\cal M}^{\rm bef}_{\gamma_T\to q_1g_2}\big|^2=\frac{e_f^2\, \alpha_{\rm em}\, \alpha_s \, 4z}{(2q^+)}
\int_{\b, \b', \r, \r'}  e^{-i\k\cdot (\b-\b')}\,  e^{i\P\cdot(\r-\r')}\int_{b^+,b^{\prime +}}\nonumber\\
&&\times\, 
\bigg\langle
\left[ {\cal U}_A^\dagger(+\infty, b^{\prime +}; \b'+z\r')\, {\cal U}_A(+\infty, b^+; \b+z\r) \right]_{b'b}
\left[ {\bar \Psi}(b^{\prime +}, \b'+z\r')t^{b'}{\cal U}_F(b^{\prime +}, -\infty;\b'+z\r'){\cal U}_F^\dagger(\b'-(1-z)\r')\right]
\nonumber\\
&&\times\, 
\gamma^-\bigg\{ \big[z^2+(1-z)^2\big]\frac{\r\cdot\r'}{|\r||\r'|}\bar Q^2K_1\big(|\r|\bar Q\big)\, K_1\big(|\r'|\bar Q\big)
+
m^2K_0\big(|\r|\bar Q\big)\, K_0\big(|\r'|\bar Q\big) 
\nonumber\\
&&
+\, 
(1-2z)\bar Q^2\, K_1\big(|\r|\bar Q\big)\, K_1\big(|\r'|\bar Q\big)\, \frac{\r^l}{|\r|}\frac{\r^{\prime l'}}{|\r'|}\frac{[\gamma^l, \gamma^{l'}]}{2}\bigg\} 
\left[ {\cal U}_F(\b-(1-z)\r){\cal U}_F^\dagger(b^+, -\infty; \b+z\r)t^b\Psi(b^+, \b+z\r)\right]
\bigg\rangle,
\ea
and
\ba
&&
 2{\rm Re} \, \frac{1}{2} \sum_{\lambda, \lambda_2h_1}\sum_{ {\rm col.}}\big( 
{\cal M}^{\rm in\dagger}_{\gamma_T\to q_1g_2} {\cal M}^{\rm bef}_{\gamma_T\to q_1g_2}
%
 \big)=-
\frac{e_f^2\, \alpha_{\rm em}\, \alpha_s\, 4z}{\big[ \P^2+(1-z)^2m^2\big]} \, \frac{(2\pi)}{(2q^+)}\, 2{\rm Re}  \int_{\b, \b', \r} e^{i\k\cdot(\b-\b')}\, e^{i\r\cdot \P}
\int_{b^+,b^{\prime +}}
\nonumber\\
&&\times 
\bigg\langle
\Big[ \bar \Psi(b^{\prime +},\b')\gamma^{-}{\cal U}_F^\dagger(+\infty, b^{\prime +}, \b')t^{a_2}\Big] 
\bigg[ iz\P^{l'}\frac{\r^l}{|\r|}{\bar Q}K_1(|\r|\bar Q)\gamma^l\gamma^{l'}+zmK_0(|\r|\bar Q)\P^l\gamma^l+i(1-z)^3m \frac{\r^l}{|\r|}\bar QK_1(|\r|\bar Q)\gamma^l\bigg]
\nonumber\\
&&
\times\, 
{\cal U}_A(+\infty, b^+;\b+z\r)_{a_2b}\, \Big[ {\cal U}_F(\b-(1-z)\r)\, {\cal U}^\dagger_F(b^+, -\infty;\b+z\r)t^b
\Psi(b^+, \b+z\r)\Big]
\bigg\rangle,
\ea
where $K_\alpha(z)$ is the modified Bessel function of the second kind and we used the notation $\bar Q^2=m^2+z(1-z)Q^2$. Moreover, the notation $\langle\cdots\rangle$ stands for  target averaging in the spirit of CGC formalism. 

\section{Production cross section in the correlation limit}

 The correlation limit, in which earlier studies have recovered the gluon TMDs from eikonal CGC calculations of $q\bar{q}$ dijets, corresponds to the limit of quasi back-to-back jets~\cite{Dominguez:2011wm}.  That limit is conveniently expressed in terms of the dijet momentum imbalance $\k$ and relative momentum $\P$, instead of the momenta $\p_1$ and $\p_2$ of the jets, which are defined as  
\be
\P=(1-z)\p_1-z\p_2 \,, \ \ \k=\p_1+\p_2
\, ,
\label{def_P_and_k}
\ee
with the lightcone momentum fraction  $z=p^+_1/(p_1^++p_2^+)$, so that $(1-z)=p^+_2/(p_1^++p_2^+) $. In particular, one has

\be
\p_1^i-\frac{p_1^+}{p_2^+}\p_2^i=\frac{\P^i}{(1-z)}
\, .
\ee
Then, Eq.~(\ref{in_S_gen}) reads
%
\ba
i{\cal M}^{\rm in}_{\gamma_T\to q_1g_2}&=& \frac{ie e_f g}{2q^+}\, \frac{(1-z)^2}
{\left[ \P^2+(1-z)^2m^2\right]}\,  {\bar u}(1)\frac{\gamma^+\gamma^-}{2} \Gamma^{\rm in}_T
\int_{\b}e^{-i\b\cdot\k}\int_{z^+} 
t^{a_2}U_F\left(+\infty,z^+;\b\right)
  \Psi(z^+,\b)
\label{in_S_corr}\, ,
\ea
%
renaming $\z$ into $\b$. 
The phase factor appearing in the before contribution (\ref{bef_S_gen}) to the S-matrix becomes  
\ba
e^{-i\v\cdot \p_1-i\z\cdot \p_2+i(\v-\z)\cdot\K}\to e^{-i\k\cdot\b+i\r \cdot(\P-\K)}
\label{phase_corr_lim}
\ea
after defining  
\ba
&&\b=z\v+(1-z)\z\, , \ \  \r=\z-\v\, .
\ea
The correlation (or back-to-back) limit is defined as $\P^2\gg \k^2$. Because of the phase factor (\ref{phase_corr_lim}), this implies 
$\r^2\ll \b^2$. In that limit, the color structure appearing in Eq.~(\ref{bef_S_gen}) simplifies as 
\ba
U_A\left(+\infty,z^+;\z\right)_{a_2b}
U_F(\v)U_F^\dagger\left(z^+,-\infty;\z\right)t^b
\Psi(z^+,\z)
&\rightarrow& U_A\left(+\infty,z^+;\b\right)_{a_2b}
U_F(\b)U_F^\dagger\left(z^+,-\infty;\b\right)t^b
\Psi(z^+,\b)\nonumber\\
&&=t^{a_2}U_F\left(+\infty,z^+;\b\right)\Psi(z^+,\b)
\label{corr_lim_Color_Bef}
\, .
\ea 
%

Using the approximation (\ref{corr_lim_Color_Bef}), the integration over $\r$ can be performed trivially and sets $\K=\P$.  All in all, the "before" contribution (\ref{bef_S_gen}) to the scattering amplitude reduces to 
\ba
i{\cal M}^{\rm bef}_{\gamma_{T,L}\to q_1g_2}&\simeq& \frac{i e e_fg}{2q^+}\frac{1}{\left[ \P^2+{\bar Q}^2\right]}\, 
{\bar u}(1)\frac{\gamma^+\gamma^-}{2} \Gamma^{\rm bef}_{T,L}
\int_{\b}e^{-i\b\cdot\k}
\int_{z^+} 
t^{a_2}U_F\left(+\infty,z^+;\b\right)\Psi(z^+,\b)
\label{bef_S_corr}
\ea
in the correlation limit $\P^2\gg\k^2$.
Note that Eqs.~(\ref{in_S_corr}) and (\ref{bef_S_corr}) contain the same $\P$-independent color structure, and differ only by a $\P$-dependent kinematical factor.

%
%

%
%

%
%
The quark-gluon dijet production cross sections are defined in Eqs. (\ref{def:xsec_L}) and (\ref{def:xsec_T}). Evaluating the cross sections in the correlation limit by using  the expressions for the scattering amplitudes given in Eqs. (\ref{in_S_corr}) and (\ref{bef_S_corr}),  one finds an expression of the form
%
\ba
(2\pi)^6 (2p_1^+) (2p_2^+)\,
\frac{d\sigma^{\gamma_{T,L}\to q_1g_2}}{dp^+_{1} d^2\p_{1}dp^+_{2} d^2\p_{2}}\bigg|_{\rm corr. lim.}=2\pi \delta(p_1^+\!+\!p_2^+\!-\!q^+) \, (4\pi)^2\alpha_{\rm em}\alpha_sC_Fe_f^2\, {\cal H}_{T,L}(\P,z,Q){\cal T}(\k)
\label{cross_sec_res_1}
\, ,
\ea
%
where the hard factors for the longitudinal and the transverse photon polarizations are\footnote{The hard factors for light dijets have been computed in the standard TMD factorization in Refs.~\cite{Boer:2016fqd,delCastillo:2021znl}. After performing the change $z\leftrightarrow (1-z)$ and setting $m=0$ in Eqs. (\ref{H_L}) and (\ref{H_T}), and noting that the different definitions of the relative transverse momentum $\P$ here and in~\cite{Boer:2016fqd,delCastillo:2021znl} coincide at leading power accuracy in the correlations limit $\P^2\gg \k^2$, one can show that the hard factors, ${\cal H}_L$ and ${\cal H}_T$, computed here match the corresponding hard factors computed in Refs.~\cite{Boer:2016fqd,delCastillo:2021znl}.} 
\ba
\label{H_L}
{\cal H}_{L}&=&  \frac{4Q^2z^3(1-z)^2}{[\P^2+{\bar Q}^2]^2}\, ,\\ 
{\cal H}_{T}&=&  z 
\left\{ \frac{(1+z^2)\P^2+(1-z)^4m^2}{[\P^2+(1-z)^2m^2]^2}+\frac{[z^2+(1-z)^2]\P^2+m^2}{[\P^2+{\bar Q}^2]^2}-\frac{2z^2\P^2}{[\P^2+{\bar Q}^2][\P^2+(1-z)^2m^2]}\right\} \, .
\label{H_T}
\ea
The target-averaged color operator ${\cal T}(\k)$ reads 
\be
{\cal T}(\k)=\int_{\b, \b'} e^{-i\k\cdot(\b-\b')}\int_{z^+,z^{\prime+}}\left\langle \bar\Psi(z^{\prime+},\b')\gamma^-U_F^\dagger(+\infty, z^{\prime+};\b')U_F(+\infty, z^+;\b)\Psi(z^+,\b)
 \right\rangle
 \label{T_op_def}
 \, .
\ee

In general, following the spirit of the CGC formalism, we have calculated the considered scattering process on a semi-classical background field representing the target, and the physical cross section is then obtained after averaging over the target background field, with a suitable probability distribution, see Eqs.~(\ref{cross_sec_res_1}) and (\ref{T_op_def}). In the context of eikonal scattering in the CGC, only the ${\cal A}^-$ component of the background is relevant, and the probability distribution is the solution of the JIMWLK evolution with, in the case of a large nucleus, initial conditions given by the MV model. By contrast, in the example considered in this study, beyond eikonal accuracy, the quark background field of the target is crucial. Two options are in principle possible. Either one could try to generalize the CGC target average, by including the quark background field as well. But no such model is available at the moment. Or one can go back to the quantum expectation value that the CGC target average is supposed to emulate. This is the approach used in this study.

For a given color operator ${\cal O}$, the CGC-like target average $\langle{\cal O} \rangle$ should be proportional to the quantum expectation value $\langle P_{{tar}}|\hat{\cal O}| P_{{tar}}\rangle$ in the state $| P_{{tar}}\rangle$ of the target with momentum $P_{{tar}}^{\mu}$. In order to ensure the proper normalization $\langle{1} \rangle =1$ and avoid ill defined expressions, one is led to the relation (see for example Refs. \cite{Belitsky:2002sm,Dominguez:2011wm,Marquet:2016cgx,Altinoluk:2019wyu})
\ba
\langle{\cal O} \rangle &=& \lim_{P_{{tar}}'\rightarrow P_{{tar}}}
\frac{\langle P_{{tar}}'|\hat{\cal O}| P_{{tar}}\rangle}{\langle P_{{tar}}'| P_{{tar}}\rangle}
\, .
\label{def_average}
\ea
We choose target states normalized as
\ba
\langle P_{{tar}}'| P_{{tar}}\rangle &=& 2P_{{tar}}^-\, (2\pi)^3 \delta(P_{{tar}}^{'-}\!-\!P_{{tar}}^-)\, \delta^{(2)}(\P_{{tar}}'\!-\!\P_{{tar}})
\, ,
\label{norm_states}
\ea
following the usual conventions from Light-Front quantization\footnote{Note that the formalism that we adopt in this paper corresponds to CGC only in a broad sense. More precisely, it corresponds to the background field method in the high-energy limit. However, we do not rely on semi-classical approximation which is crucial for the CGC effective theory in the strict sense. Eq. (\ref{def_average}) is then the relation between the quantities calculated in the background field method and their counter parts in the full quantum field theory.}. In general, in quantum field theory, the translation of any local operator $\hat{\cal O}(x)$ is obtained by the action of the momentum operator $\hat P_\mu$ as
\ba
\hat{\cal O}(x)=e^{ia^\mu\hat P_\mu}\hat{\cal O}(x-a)e^{-ia^\mu\hat P_\mu}
\, .
\ea
Hence, the matrix elements of non-local operators behave as 
\ba
\langle P_{{tar}}'| \hat{\cal O}_1(x_1)\cdots \hat{\cal O}_n(x_n) | P_{{tar}}\rangle&=&
\langle P_{{tar}}'| e^{ia^\mu\hat P_\mu}\hat{\cal O}_1(x_1-a)\cdots \hat{\cal O}_n(x_n-a)e^{-ia^\mu\hat P_\mu} | P_{{tar}}\rangle\nonumber\\
&=& e^{ia^{\mu}[(P_{{tar}}')_{\mu}-(P_{{tar}})_\mu]}
\langle P_{{tar}}'| \hat{\cal O}_1(x_1-a)\cdots \hat{\cal O}_n(x_n-a) | P_{{tar}}\rangle
\label{transl_correlator}
\ea
under global translations. 

Using the relations (\ref{def_average}),   (\ref{norm_states}) and (\ref{transl_correlator}), one can calculate target-averaged color operator ${\cal T}(\k)$ 
from Eq. (\ref{T_op_def}) as 
\ba
{\cal T}(\k)&=&\lim_{P_{{tar}}'\rightarrow P_{{tar}}}
\int_{\b, \b'} \frac{e^{-i\k\cdot(\b-\b')}}{\langle P_{{tar}}'| P_{{tar}}\rangle}\int_{z^+,z^{\prime+}}\left\langle P_{{tar}}'\left|  \bar\Psi(z^{\prime+},\b')\gamma^-U_F^\dagger(+\infty, z^{\prime+};\b')U_F(+\infty, z^+;\b)\Psi(z^+,\b)
\right| P_{{tar}} \right\rangle
\nonumber\\
&=&\lim_{P_{{tar}}'\rightarrow P_{{tar}}}
\int_{\b, \b'} \frac{e^{-i\k\cdot(\b-\b')}}{\langle P_{{tar}}'| P_{{tar}}\rangle}\int_{z^+,z^{\prime+}}
e^{iz^+(P_{{tar}}^{'-}\!-\!P_{{tar}}^-)}\, e^{-i\b\cdot(\P_{{tar}}'\!-\!\P_{{tar}})}
\nonumber\\
&&\hspace{2cm} \times\,
\left\langle P_{{tar}}'\left|  \bar\Psi(z^{\prime+}\!-\!z^+,\b'\!-\!\b)\gamma^-U_F^\dagger(+\infty, z^{\prime+}\!-\!z^+;\b'\!-\!\b)U_F(+\infty, 0;{\bf 0})\Psi(0,{\bf 0})
\right| P_{{tar}} \right\rangle
 \nonumber\\
&=&\lim_{P_{{tar}}'\rightarrow P_{{tar}}}
\int_{\Delta\b} \frac{e^{i\k\cdot\Delta\b}}{2P_{{tar}}^-}\int_{\Delta z^+}
\left\langle P_{{tar}}'\left|  \bar\Psi(\Delta z^+,\Delta\b)\gamma^-U_F^\dagger(+\infty, \Delta z^+;\Delta\b)U_F(+\infty, 0;{\bf 0})\Psi(0,{\bf 0})
\right| P_{{tar}} \right\rangle
 \label{T_op_result}
 \, ,
\ea
in which the limit $P_{{tar}}'\rightarrow P_{{tar}}$ can now be taken safely.


The unpolarized TMD quark distribution is defined (up to UV and rapidity regularization issues) as 
\be
f_1^q({\rm x},\k)=\frac{1}{(2\pi)^3}\int_{\b} e^{i\k\cdot\b}\int_{z^+}e^{-iz^+{\rm x}P_{{tar}}^-}\langle P_{{tar}}|\bar \Psi(z^+,\b)\frac{\gamma^-}{2}U_F^\dagger(+\infty, z^+; \b)U_F(+\infty, 0; {\bf 0})\Psi(0,{\bf 0})|P_{{tar}}\rangle,
\label{TMD_def}
\ee
where we have neglected the transverse gauge link at infinity. 
Comparing Eqs. (\ref{T_op_result}) and (\ref{TMD_def}), one finds the relation
%
\be
{\cal T}(\k)=\frac{(2\pi)^3}{P_{{tar}}^-}f^q_1({\rm x}=0,\k).
\label{rel_T_op_vs_TMD}
\ee
Hence, the differential cross section (\ref{cross_sec_res_1}) in the correlation limit becomes a factorization formula involving the quark TMD distribution at ${\rm x}=0$ target momentum fraction:
\ba
 p_1^+\, p_2^+\,
\frac{d\sigma^{\gamma_{T,L}\to q_1g_2}}{dp^+_{1} d^2\p_{1}dp^+_{2} d^2\p_{2}}\bigg|_{\rm corr. lim.}=(2q^+)\, \delta(p_1^+\!+\!p_2^+\!-\!q^+) \, \frac{\alpha_{\rm em}\alpha_sC_Fe_f^2}{W^2} \, {\cal H}_{T,L}(\P,z,Q)\, f^q_1({\rm x}=0,\k)
\label{cross_sec_res_1bis}
\, .
\ea
Note that in the denominator of Eq.~(\ref{cross_sec_res_1bis}) we have made the high-energy approximation $ W^2=(q+P_{tar})^2\simeq 2q^+P_{{tar}}^-$ with $W$ being the center of mass energy of the photon-target collision. This suppression of Eq.~(\ref{cross_sec_res_1bis}) as $1/W^2$ at high energy is characteristic of a next-to-eikonal contribution, and here it is due to the exchange of quarks in the $t$-channel instead of only gluons.

\begin{figure}[htb]
 \includegraphics[width=10cm]{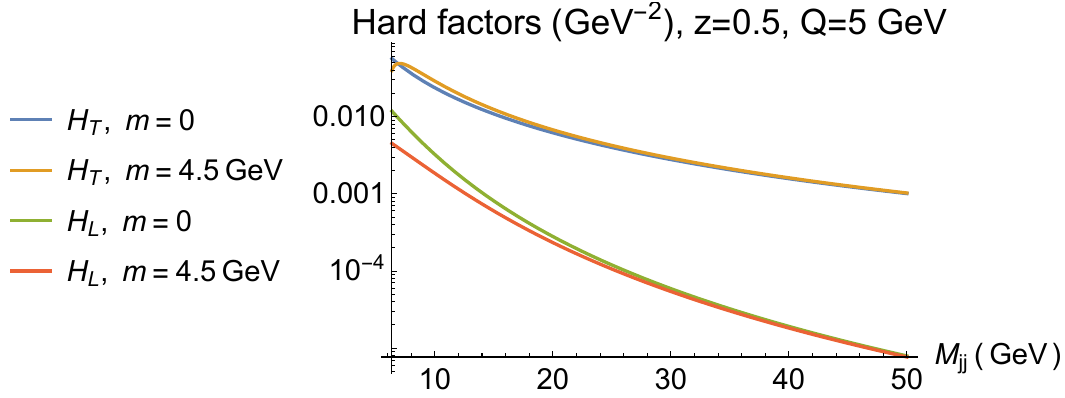}
 \vskip 0.8cm
 \includegraphics[width=10cm]{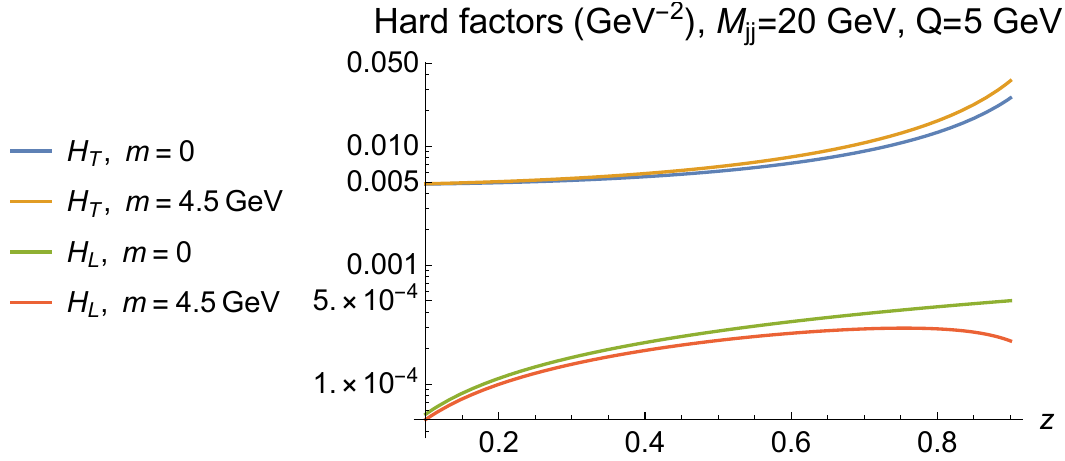}
 \vskip 0.8cm
\includegraphics[width=10cm]{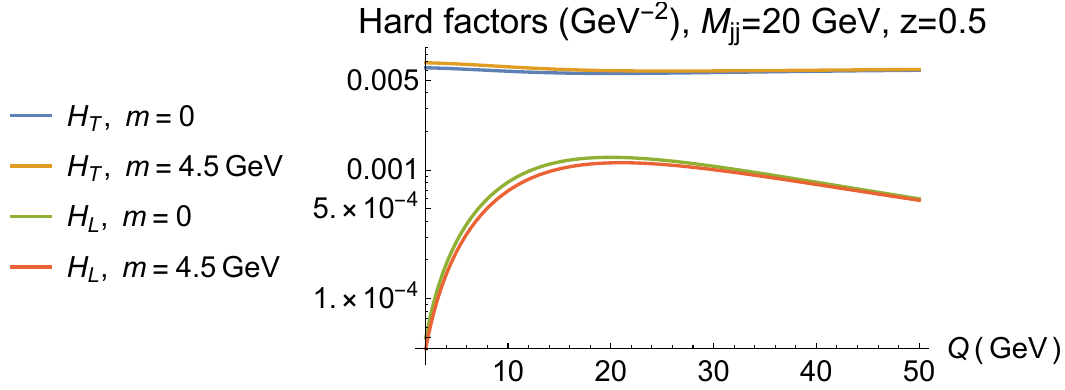}
\caption{Hard factors, (\ref{eq:HL}) and (\ref{eq:HT}), versus $M_{jj}$ (top), $z$ (middle, for values such that $|\P|>6$ GeV) and $Q$ (bottom), for massless and massive ($m=4.5$ GeV) quarks.}
\label{fig:hardfactors}
\end{figure}

 Alternatively, in the correlation limit we can write the cross section in terms of $\k$, $z$ and the dijet mass $M_{jj}$ defined as
%
%
$M^2_{jj}\equiv(p_1+p_2)^2={\P^2}/{(z(1-z))}+{m^2}/{z}$, and obtain
\be
\frac{d\sigma^{\gamma_{T,L}\to q_1g_2}}{dz\, dM^2_{jj}\,  d^2\k }\bigg|_{\rm corr. lim.}=\, (2\pi)\, \frac{\alpha_{\rm em}\, \alpha_s\, C_F\, e_f^2}{W^2}\, \widetilde{\cal H}_{T,L}(M_{jj},z,Q)f_1^q({\rm x}=0, \k)
\,  ,
\label{eq:corrlim2}
\ee
with the new hard factors
\be
 \widetilde{\cal H}_{L}(M_{jj},z,Q)=\frac{4Q^2\, z(1-z)^2}{\left[ (1-z)(M^2_{jj}+Q^2)+m^2\right]^2}\, ,
  \label{eq:HL}
 \ee
%
%
\ba
 \widetilde{\cal H}_{T}(M_{jj},z,Q)
 &=&
 \frac{(1+z^2)}{(1-z)}\frac{1}{\left[M^2_{jj}-m^2\right]}
+ \frac{(1-2z)}{\left[ (1-z)(M_{jj}^2+Q^2)+m^2\right]}
 +\frac{(1-z)\left[2m^2-\big(z^2+(1-z)^2\big)Q^2\right]}{\left[ (1-z)(M_{jj}^2+Q^2)+m^2\right]^2}\nonumber\\
 &&-\frac{2m^2}{\left[M^2_{jj}-m^2\right]^2}
 +\frac{2z(1-z)m^2}{\left[M^2_{jj}-m^2\right]\left[ (1-z)(M_{jj}^2+Q^2)+m^2\right]}
 \label{eq:HT}
 \, ,
 \ea
illustrated on Fig.~\ref{fig:hardfactors} for massless and bottom quarks. For the chosen values of the variables, 
$\widetilde{\cal H}_{T}$ is much larger than $\widetilde{\cal H}_{L}$, hard factors decrease  with $M_{jj}$, and $\widetilde{\cal H}_{T}$ is notably flat with $Q$. 
Overall, the influence of the quark mass $m$ is moderate, except in the small $M_{jj}$ region close to the kinematical limit, and for $z\to 1$. However, these two limits are outside of the validity region of our approach. 
%


 %

\section{Discussion}

In this paper we have studied the process of quark-gluon dijet production in DIS at high energy in the CGC. Such process is mediated by a quark exchange in the $t$-channel and thus it is sub-eikonal. Going to the correlation limit of almost back-to-back jets, we have established the relation with the unpolarized quark TMD distribution at small $x$.

At this stage, we would like to comment on the difference between the quark TMD appearing in our results and the one appearing in Ref. \cite{Marquet:2009ca}.
In the present manuscript, the operator definition of the quark TMD (up to regularization issues) is obtained from the analysis of the cross section and therefore it is the full quark TMD distribution containing both sea and valence contributions. By contrast, in Ref. \cite{Marquet:2009ca}, the authors calculate SIDIS in the dipole approach, in pure gluon background field, and interpret part of their result as the sea contribution to the quark TMD. This is analog to write, in the collinear factorization, the sea contribution to the quark PDF as the gluon PDF convoluted with a coefficient accounting for a gluon to quark extra step of DGLAP evolution.

The determination of the quark TMD distributions proposed here relies on the identification of quark-gluon jets in DIS. In the case of light quarks, such identification is challenging, see e.g. \cite{Larkoski:2017jix,Kogler:2018hem} and refs. therein. However, machine learning techniques might improve the discrimination of light quarks and gluon jets in the future \cite{Lee:2022kdn}. 

Another possibility would be considering heavy flavor jets, whose discrimination from light quark or gluon jets is better established. The expressions provided in this paper consider quark masses so they can be applied for heavy flavors. Of course, that would restrict the study to heavy flavor TMD distributions in the target.  

On the other hand, the process studied here competes with the eikonal production of a quark, an antiquark and a gluon at partonic level, where the quark and gluon fragment into separate jets and the antiquark does not belong to these two tagged jets. Even though this process is of higher order in $\alpha_s$ compared to the mechanism studied in this manuscript, in the high energy limit it is enhanced since it is an eikonal contribution and not a next-to-eikonal one. A proper comparison of both mechanisms in the EIC kinematics requires a dedicated effort and it is beyond the scope of this work. Nevertheless, it is expected that the contamination by this background mechanism can be significantly reduced by imposing appropriate experimental cuts.

Indeed, in this background mechanism, the antiquark is typically carrying away sizable longitudinal and transverse momenta. 
The contribution (\ref{eq:corrlim2}) obeys $p^+_1+p_2^+=q^+$ exactly, whereas the background contribution obeys $0<p^+_1+p_2^+<q^+$, with no expected enhancement for $p^+_1+p_2^+ \to q^+$.
Hence, experimentally, one can impose that the total lightcone momentum of the dijet $p^+_1+p_2^+$ is close enough to the incoming photon momentum $q^+$ in order to significantly reduce the background contribution.
Similarly, in the background contribution the transverse momentum of the produced antiquark is typically of the same order as the one of the produced quark. For that reason, imposing the back-to-back limit $\P^2\gg \k^2$ with $\P$ and $\k$ defined as in Eq. (\ref{def_P_and_k}) from the quark and gluon jets momenta should further suppress the background contribution.
While additional studies are required, we think that with these experimental cuts the process of quark-gluon dijet production in DIS at the EIC is a promising new method to probe the quark TMD distribution in addition to semi-inclusive DIS and Drell-Yan production, see e.g.~\cite{Boussarie:2023izj,Angeles-Martinez:2015sea} and refs. therein.

\begin{acknowledgments}
NA has received financial support
from Xunta de Galicia (Centro singular de investigaci\'on de Galicia
accreditation 2019-2022), by European Union ERDF, and by the
Spanish Research State Agency under project PID2020-119632GB-I00. GB is supported in part by the National Science Centre (Poland) under
the research grant no. 2020/38/E/ST2/00122 (SONATA BIS 10). This work has been performed in
the framework of the European Research Council project ERC-2018-
ADG-835105 YoctoLHC and the MSCA RISE 823947 ``Heavy ion collisions: collectivity and precision in saturation physics" (HIEIC), and
has received funding from the European Union's Horizon 2020 research
and innovation programme under grant agreement No. 824093.
\end{acknowledgments}

\bibliography{mybib}

\end{document}